\documentclass[12pt,reqno]{article}

\usepackage{amsmath, amsthm, amssymb,epsfig,alltt}

\def\a{\alpha}
\def\b{\beta}
\def\bt{{\beta_T}}

\def\e{\epsilon}
\def\p{\partial}

\def\t{\tau}
\def\th{\theta}
\def\s{\sigma}

\def\o{\omega}

\def\whata{{\widehat{\alpha}}}
\def\whatb{{\widehat{\beta}}}

\def\nn{\nonumber}

\def\2pap{2\pi\alpha^\prime}

\def\beq{\begin{eqnarray}}
 \def\eeq{\end{eqnarray}}
 \def\4pap{4\pi\a^\prime}

 \def\ap{{\a^\prime}}

 \def\ta{{\tilde \a}}
 \def\tb{{\tilde \b}}

 \def\tal{{\tilde \alpha}}
 \def\tbe{{\tilde \beta}}

 \def\bole{{\boldsymbol \epsilon}}
 \def\bolth{{\boldsymbol \theta}}

 \def\bolk{{\boldsymbol k}}
 \def\bolX{{\boldsymbol X}}
 \def\calO{{\cal O}}

\begin{document}


\title{\bf Exact Duality of \\
The Dissipative Hofstadter Model \\
on a Triangular Lattice}

\author{Taejin Lee \\~~\\
Department of Physics, Kangwon National University, \\
Chuncheon 200-701 Korea }

\maketitle

\centerline{\bf Astract}
We study the dissipative Hofstadter model on a triangular lattice, making use of the $O(2,2;R)$ 
T-dual transformation of string theory. The $O(2,2;R)$ dual transformation transcribes the model 
in a commutative 
basis into the model in a non-commutative basis. 
In the zero temperature limit, the model exhibits an exact 
duality, which identifies equivalent points on the two dimensional parameter space of the model. 
The exact duality 
also defines magic circles on the parameter space, where the model can be mapped onto the boundary 
sine-Gordon on a triangular lattice. The model describes the junction of three quantum wires 
in a uniform magnetic field background. An explicit expression of the equivalence relation,
which identifies the points on the two dimensional parameter space of the model by the 
exact duality, is obtained. It may help us to understand the structure of the phase diagram of the model. 



\vskip 2cm

\section{INTRODUCTION}
    
The dualities and critical behaviors of the low dimensional quantum systems are fascinating 
subjects to explore, as many recent discoveries in string theory and condensed matter 
physics are based on them. Among others, good examples include the rolling tachyons 
\cite{Sen:2002nu, senreview}, the target space duality \cite{tdual}
and the non-commutative geometry \cite{seib} in string theory 
and the Tomonaga-Luttinger liquid \cite{Tomonaga,Luttinger} with impurity \cite{Kane:1992b, Kane:1992a}, 
the Kondo problem \cite{Kondo}, and the junctions of quantum wires \cite{chamon,oshikawa2006} 
in condensed matter physics. The dissipative Hofstadter model 
on a triangular lattice, which we will discuss in the present work, would serve also as an excellent 
example. The Hofstadter model, which is also known as Wannier-Azbel-Hofstadter model \cite{Azbel,Hofs,Wannier}, describes quantum particles moving in two dimensions, subject to a uniform magnetic field and a periodic potential, 
has been extensively studied for many decades as a quantum mechanical model of the quantum Hall effect \cite{Klitzing, Laughlin, Thouless}. 
If the frictional force of Caldeira-Legget type \cite{caldeira83ann, caldeira83phy} 
is introduced to the Hofstadter model, the phase diagram of the model becomes even more complex, yet more interesting. This model is called the dissipative Hofstadter model
(DHM) \cite{callan91, Callan:1992vy, freed93, Callan:1995ck}, which appears in disguise 
in many places of theoretical physics. The friction force of 
Caldeira-Legget type
is produced by coupling the quantum particles to an infinite number of harmonic oscillators, which 
depict degrees of freedom of the environment or the bath. In quantum theory the coupling to the bath 
produces a non-local effective interaction, which can be traded with the local Polyakov term in string theory.
It enables us to have a string theory representation of the DHM. In string theory the model can be understood as a model of open string in the background of the Neveu-Schwarz (NS) B-field with a periodic potential at the 
ends of the open string, which may be realized as the tachyon condensation.

Once reformulating the DHM as a string theory, we can take advantage of the recent developments in string theory, such as the target space duality (T-duality) \cite{tdual} and the non-commutative geometry \cite{seib}. 
The exact duality of the DHM on a square lattice \cite{callan91} has been identified 
as a subgroup of the T-dual symmetry group in string theory \cite{Lee2009}, unbroken in the zero temperature by the periodic potential. The particle-kink duality of the DHM model, which was called 
previously the approximate duality \cite{callan91}, also has been shown to hold exactly \cite{Lee2009}
in the framework of string theory, regardless of the strength of the magnetic field. 

The DHM makes its appearance also
in the quantum impurity problems \cite{ Kane:1992b,  Kane:1992a, fisher;1985, furusaki} 
in one dimensional condensed matter 
system. If the magnetic field is turned off, the model reduces to the Schmid model 
\cite{schmid, guinea} (or the boundary sine-Gordon model), which consists of one 
dimensional Tomonaga-Luttinger (TL) liquid on a half line and a boundary periodic potential at the origin as an interaction term with an impurity. The string coordinate field corresponds to the bosonized field of the
TL fermion field on each lead and the Regge slope $\ap$ of string theory is identical to the inverse of the TL 
parameter.

The junctions of quantum wires \cite{chamon,oshikawa2006,larkin,fazio, sodano}
are also places where the DHM plays an important role. In the 
absence of the magnetic field, the TL liquid on the quantum wires is described by the free string action 
and the electron transport between wires may be represented in the bosonized theory by the boundary 
periodic potential. The DHM with the magnetic field may serve as a model of the junction of quantum wires
enclosing a magnetic flux. 

Although the DHM has been studied in connection with diverse subjects in theoretical physics, 
most of the studies have been confined to the case of the periodic potential on a square lattice. 
With the DHM model on a square lattice, we are only able to describe the junction of two wires.
In order to study the critical behaviors of the junction of three quantum wires, which 
is the basic building block of the circuits made of quantum wires, we need to extend it to the DHM model on a 
triangular lattice.

\section{The Dissipative Hofstadter model on a Triangular Lattice }

The dissipative Wannier-Azbel-Hofstadter model on a triangular lattice is described by the following 
action 
\beq  \label{dhmaction}
S &=& \frac{\eta}{4\pi \hbar} \int^{\bt/2}_{-\bt/2} dt dt^\prime 
\frac{\left(\bolX(t) - \bolX(t^\prime)\right)^2}{(t-t^\prime)^2} + 
\frac{ieB_H}{2\hbar c} \int^{\bt/2}_{-\bt/2} dt \sum_{a,b=1}^2\epsilon^{ab}
\p_t X^a X^b \nn\\
&& + \frac{V_0}{\hbar} \int^{\bt/2}_{-\bt/2} dt \sum_{a=1}^2
\cos \frac{2\pi \bolk^a \cdot \bolX}{l},
\eeq
where $\b_T = 1/T$ and 
\beq \label{lattice}
\bolk_1 = (\frac{1}{2},\frac{\sqrt{3}}{2}), ~~~
\bolk_2 = (\frac{1}{2},-\frac{\sqrt{3}}{2}), ~~~
\bolk_3 = \left(-1,0 \right).
\eeq
The first term is the effective non-local action of  Caldeira-Legget type
dissipation, obtained by integrating out the bath degrees of freedom, represented by 
an infinite number of harmonic oscillators. The strength of the coupling between the 
quantum particle and the bath is measured by the frictional constant $\eta$. The second term 
denotes the interaction with the uniform magnetic field $B_H$. The third term is the periodic 
potential on the triangular lattice.

Scaling the coordinate fields $\bolX$ and defining the world sheet parameter $\s$ as 
follows 
\beq
\bolX \rightarrow \frac{l}{2\pi} \bolX, ~~~ \s = 2\pi \b_T t, 
\eeq
we can map the DHM action Eq.(\ref{dhmaction}) onto the string theory action on a 
cylindrical surface in the background of the 
uniform NS B-field with a periodic potential 
\beq
S&=&  \frac{1}{4\pi}
\int d\t d\s \sum_{a, b =1}^2 E_{ab} \left(\p_\t + i\p_\s\right) X^a 
\left(\p_\t -i \p_\s\right) X^b\nn\\
&&  + \frac{V_0}{2} \int d\s \sum_{a=1}^3 \left(e^{i\bolk^a\cdot \bolX}+ 
e^{-i\bolk^a\cdot \bolX}\right) \label{stringaction}
\eeq 
where $E_{ab} = \a\delta_{ab} + 2\pi B_{ab}= \a\delta_{ab} + \b \epsilon_{ab}$, 
and $2\pi\b = \frac{eB_H}{\hbar c}l^2$. The parameter 
$\a$ is the TL parameter, which is related to the Regge slope in string theory and the 
friction constant as 
\beq
\a = 1/\ap=\eta/2\pi .
\eeq

The same periodic potential in the string theory action Eq.(\ref{stringaction}) may arise
in the model of junction of three quantum wires \cite{oshikawa2006}. 
Let us denote the TL fermion field
on each quantum wire as $\psi^a_{L/R}$, $a= 1, 2, 3$. Then the hopping interaction, which 
is responsible for the electron transport between the wires may be written as 
\beq
\sum_{a=1}^3 \left(\psi^{a\dag}_L \psi_L^{a+1}- \psi^{a\dag}_R \psi^{a+1}_R   \right)
\eeq
where $\psi^{4}_{L/R} = \psi^1_{L/R}$. Making use of the Fermi-Bose equivalence \cite{Lee2009q,Leeklein}
\begin{subequations}
\beq
\psi^1_L &=& e^{-\frac{\pi}{2} i \left(p^1_L + p^1_R  \right)} e^{-\sqrt{2} i \phi^1_L}, \label{n3fermion1}\\
\psi^2_L &=& e^{-\frac{\pi}{2} i \left(p^2_L+ 2 p^1_L + p^2_R + 2p^1_R \right)} e^{-\sqrt{2} i \phi^2_L}, \label{n3fermion2}\\
\psi^3_L &=& e^{-\frac{\pi}{2} i \left(p^3_L +2 p^2_L + 2p^1_L + p^3_R + 2p^2_R+ 2p^1_R\right)} e^{-\sqrt{2} i \phi^3_L}, \label{n3fermion3}\\
\psi^1_R &=& e^{-\frac{\pi}{2} i \left(p^1_L + p^1_R  \right)} e^{\sqrt{2} i \phi^1_R}. \label{n3fermion5}, \\
\psi^2_R &=& e^{-\frac{\pi}{2} i \left(p^2_L+2 p^1_L+ p^2_R+ 2p^1_R\right)} e^{\sqrt{2} i \phi^2_R}, \label{n3fermion4}\\
\psi^3_R &=& e^{-\frac{\pi}{2} i \left(p^3_L +2 p^2_L+ 2p^1_L + p^3_R + 2p^2_R+ 2p^1_R\right)} e^{\sqrt{2} i \phi^3_R}, \label{n3fermion6}
\eeq
\end{subequations}
and the Neumann condition in the fermion theory 
\beq
\psi^a_L \vert {\bf N} \rangle = i \psi^{a\dag}_R \vert {\bf N} \rangle, ~~~
\psi^{a\dag}_L \vert {\bf N} \rangle = i \psi^{a}_R \vert {\bf N} \rangle,~~~ a = 1, 2, 3,
\eeq
or the Neumann condition in the boson theory
\beq
\phi^a_L \vert {\bf N} \rangle = \phi^a_R \vert {\bf N} \rangle, ~~~ a =1, 2, 3 , 
\eeq
we may rewrite the hopping interaction between the quantum wires in the bosonized theory as 
\beq
\sum_{a=1}^3 \left( e^{i \frac{\phi^a-\phi^{a+1}}{\sqrt{2}}}+ 
e^{-i \frac{\phi^a-\phi^{a+1}}{\sqrt{2}}} \right),
\eeq
where $\phi^4 = \phi^1$ and $\phi^a = \phi^a_L + \phi^a_R$. It is noteworthy that non-trivial 
Klein factors do not appear in the interaction of boson form if the Klein factors for the 
fermion fields are chosen judiciously.

Applying an $SO(3)$ rotation to the boson fields $(\phi^1, \phi^2, \phi^3)$,
\begin{subequations}
\beq
\phi^1 &=& \frac{1}{\sqrt{2}} X^1 
+ \frac{1}{\sqrt{6}} X^2 + \frac{1}{\sqrt{3}} X^3, \\
\phi^2 &=& -\frac{1}{\sqrt{2}} X^1 
+ \frac{1}{\sqrt{6}} X^2 + \frac{1}{\sqrt{3}} X^3, \\
\phi^3 &=& -\sqrt{\frac{2}{3}} \,\,X^2 + \frac{1}{\sqrt{3}} X^3, 
\eeq
\end{subequations}
brings us to the periodic potential on a triangular lattice Eq.(\ref{stringaction})
\beq
\sum_{a=1}^3 \left(e^{i\bolk^a\cdot \bolX}+ 
e^{-i\bolk^a\cdot \bolX}\right).
\eeq
Note that the third string coordinate field $X^3$ does not appear in the periodic potential. 
It is an auxiliary field. Thus, the junction of three quantum wires is described by the 
DHM on a triangular lattice. 

\section{The Target Space Dual Transformation: $O(2,2;{\bf R})$}

The string coordinate fields $X^a$, $a= 1, 2$ may be expanded in terms of the normal mode operators 
at the boundary ($\t=0$) as
\begin{subequations} 
\beq
X^a &=& X^a_L + X^a_R, \\
X^a_L &=& \frac{1}{\sqrt{2}} x^a_L+ \frac{1}{\sqrt{2}} p^a_L \sigma+\frac{i}{\sqrt{2}}
\sum_{n\neq 0}\frac{\a^a_n}{n}e^{-ni\sigma},  \\
X^a_R &=& \frac{1}{\sqrt{2}} x^a_R- \frac{1}{\sqrt{2}} p^a_R \sigma+\frac{i}{\sqrt{2}}\sum_{n\neq 0}\frac{\tilde\a^a_n}{n}e^{ni\sigma},
\eeq
\end{subequations}
where the normal mode operators satisfy the canonical commutation relations as
\beq
\left[ x^a,p^b\right]&=&i\delta^{ab},~
\left[ \a^a_m, \a^b_n \right] = g^{ab}m\delta_{{m+n},0},\nn\\ 
\left[\tilde\a^a_m, \tilde\a^b_n \right] &=& g^{ab} m\delta_{{m+n},0}, \quad g^{ab} = \a^{-1}\delta^{ab}.
\eeq
In the absence of the periodic potential the string state at $\t=0$, satisfies the following boundary 
condition, which is expressed as a boundary condition for the boundary state $\vert B_E \rangle$ 
\beq
\left(E_{ab} \a^b_{-n} + E^t_{ab} \ta^b_n\right) |B_E\rangle= 0, 
\quad p^b |B_E\rangle= 0, ~~ a, b = 1, 2 .
\eeq
If the magnetic field is turned off, the boundary condition for $|B_E \rangle$
reduces to the Neumann condition
\beq
\left(\a^a_{-n} + \ta^a_n\right) |B_E\rangle= 0, 
\quad p^a |B_E\rangle= 0, ~~ a = 1, 2 .
\eeq

If we turn off the boundary periodic potential, the string theory action 
Eq.(\ref{stringaction}) reduces to the action of closed string in the background of NS B-field, which
is invariant under $O(2,2;{\bf R})$ T-dual transformation \cite{tdual}
\beq \label{tdualz}
E \rightarrow
\bar{E} = (a E + b)(c E + d)^{-1}
\eeq
where $a,\, b,\, c$ and $d$ satisfy the $O(2,2;{\bf R})$ condition
\beq
\left(\begin{array}{cc} 
  a & b \\
  c & d  
\end{array}\right)^t 
\left(\begin{array}{cc}
  0 & I \\
  I & 0 
\end{array}\right) 
\left(\begin{array}{cc}
  a & b \\
  c & d  
\end{array}\right) &=& 
\left(\begin{array}{cc}
  0 & I \\
  I & 0  
\end{array}\right).
\eeq
Under the T-dual transformation Eq.(\ref{tdualz}) the left and right movers 
transform as
\beq
\a_n(E)  \rightarrow  (d- cE^t)^{-1} \a_n(\bar E),  ~~
\tilde{\a}_n(E) \rightarrow (d+ cE)^{-1}\tilde{\a}_n(\bar E).
\eeq
By this $O(2,2;{\bf R})$ T-dual transformation, the boundary condition for $|B_E\rangle$ 
can be transcribed into 
the usual Neumann condition in a new oscillator basis 
$\{\beta^a_n, \tilde\b^a_n\}$ as
\beq \label{neumann}
\left(\b^a_{-n} + \tilde \b^a_{n}\right) |B_E\rangle = 0, ~~~ a= 1, 2 .
\eeq

It has been shown that 
two bases $\{\a^a_n, \tilde\a^a_n; a =1, 2, n\in {\bf Z} \}$ and 
$\{\beta^a_n, \tilde\b^a_n; a= 1, 2, n \in {\bf Z}\}$ are related to 
each other by a $O(2,2;{\bf R})$ T-dual transformation generated by $T$ in ref.
\cite{Lee2009,Lee;canonical}
\begin{subequations}
\beq 
T &=& \left(\begin{array}{cc} I & 0 \\ \bolth/(2\pi) & I \end{array} \right),\label{t1}\\
\bolth/(2\pi) &=& \frac{1}{E} (2\pi B) \frac{1}{E^t} = \frac{\b}{\a^2+\b^2} \bole,\label{t2}\\
\a^a_n &=& \left(G(E)^{-1}\right)^a{}_b \,\b^b_n,\\
\tal^a_n &=& \left(G(E^t)^{-1}\right)^a{}_b \,\tbe^b_n
\eeq
\end{subequations}
where 
\beq
G= E^t g^{-1} E = \left(\frac{\a^2+\b^2}{\a}\right) \, I.
\eeq
It should be noted that the oscillators $\{\beta^a_n, \tilde\b^a_n; a= 1, 2, \,n \in {\bf Z}\}$
respect the worldsheet metric $G$
\beq
[\b^a_n, \b^b_m ] = (G^{-1})^{ab} n 
\delta(n+m), ~~~
[ \tbe^a_n, \tbe^b_m ] =&(G^{-1})^{ab} n 
\delta(n+m) 
\eeq
and
the string coordinate operators $X^a$, $a=1, 2$ are no longer commuting operators in the new basis
\cite{lee002,lee003,lee004} in the zero temperature limit where $\b_T \rightarrow \infty$
\beq
\left[X^a(\s_1), X^b (\s_2) \right] &=& i\, \th^{ab}. \label{noncommutativity}
\eeq
This is precisely the non-commutative relation between the open string coordinate operators 
\cite{seib,Lee;open,lee0105}.
It is the closed string theory realization of the non-commutativity which is mainly discussed in the 
context of the open string theory.
In the open string theory the algebra of the coordinate operators, defined 
at equal $\t$ at end points is non-commutative. In closed string theory, as the world sheet parameters
are interchanged, these points are on the boundary $\t=0$ at equal $\s$. Thus, the non-commutative
algebra of open string is expected to emerge in the low temperature limit or the equal $\s$ limit in the 
closed string theory \cite{Leequon}.

In the new oscillator basis the coordinate operators $X^a(\s,0)$ $a=1, 2$ at the 
boundary may be written as 
\begin{subequations}
\beq
X^a(\s,0) &=& Z^a(\s,0) + \frac{i}{\sqrt{2}} \frac{\b}{\a} \sum_{n\not=0} \frac{1}{n}
\e^{ab} \left(\b^b_n+\tb^b_{-n} \right) e^{in\s}, \label{decomp1}\\
Z^a(\s,0) &=& x^a + \o^a \s + i\frac{1}{\sqrt{2}} \sum_{n\not=0} \frac{1}{n}
\left[\b^a_n e^{in\s} + \tb^a_n e^{-in\s}\right], \label{decomp2}
\eeq
\end{subequations}
where $\o^a \in {\bf Z},~ a= 1, 2$, are winding numbers.
Here $Z^a$, $a=1,2$, are commuting coordinate operators of the closed string with the world sheet metric 
$G_{ab}$. This decomposition is useful when we evaluate the boundary state and the partition function.

\section{Boundary State and Magic Circles}

The boundary state formulation \cite{callan90} is one of the most efficient methods to evaluate 
the partition function and the correlation functions of operators. 
The partition function and the correlation functions of the operators $\calO_i$, $i = 1, \dots, n$, 
are calculated in the boundary state formulation as 
\begin{subequations}
\beq
Z &=& \langle 0 \vert B \rangle, \\
\langle T \calO_1 \dots \calO_n \rangle &=& \langle 0 \vert :  \calO_1 \dots \calO_n: \vert B \rangle . \label{correlation}
\eeq
\end{subequations}
The boundary state corresponding to the DHM on a triangular lattice may be written as 
\beq 
|B\rangle &=& {\boldsymbol T}\,\exp\Biggl[\frac{V_0}{2} \int d\s
\sum_{a=1}^3 \left(e^{i\bolk^a\cdot \bolX}+ e^{-i\bolk^a\cdot \bolX}\right)\Biggl\vert_{\t=0}
\Biggr]|B_E\rangle . \label{boundary}
\eeq 
Here ${\boldsymbol T}$ is the $\s$-odering, which is equivalent to the time ordering. (Recall that the 
Euclidean time $t$ is replaced by the world sheet coordinate $\s$). 

We may rewrite the periodic potential term as follows 
\beq
\sum_{a=1}^3 \left(e^{i\bolk^a\cdot \bolX}+ e^{-i\bolk^a\cdot \bolX}\right) &=& 
\sum_{a=1}^3 \Biggl\{ \exp \left(i\sum_{b=1}^2 \sqrt{\frac{3}{2}} R_{ab} X^b \right) \nn\\
&& +  
\exp \left(-i\sum_{b=1}^2 \sqrt{\frac{3}{2}} R_{ab} X^b \right)
\Biggr\}
\eeq
where $R_{ab}$, for $a = 1, 2, 3$ and $b= 1, 2$, are the components of $3 \times 2$ submatrix of 
an $SO(3)$ rotation matrix $(R)$
\beq
(R) =\left(\begin{array}{rrr} \frac{1}{\sqrt{6}} & \frac{1}{\sqrt{2}} & \frac{1}{\sqrt{3}} \\
-\frac{1}{\sqrt{6}} & \frac{1}{\sqrt{2}} & -\frac{1}{\sqrt{3}} \\ 
-\frac{\sqrt{2}}{\sqrt{3}} & 0 & \frac{1}{\sqrt{3}} \end{array} \right), ~~~ (R)^t (R) = (R) (R)^t = I.
\eeq

If we expand the boundary state in $V_0$, we find 
\beq
\vert B \rangle &=& \sum_{n^1, n^2, n^3} \frac{1}{n^1! n^2! n^3!} 
\left(\frac{V_0}{2} \right)^{n^1+ n^2 +n^3} \int \prod^{n^1}_{i=1} d\s^1{}_i
\prod^{n^2}_{i=1} d\s^2{}_i\prod^{n^3}_{i=1} d\s^3{}_i \nn\\
&& ~~~~{\boldsymbol T}\exp \left\{ i \sqrt{\frac{3}{2}} \sum_{a=1}^3 \sum_{b=1}^2\sum_{i=1}^{n^a} e^a{}_i R_{ab} X^b (\s^a{}_i) 
\right\} \vert B_E \rangle
\eeq
where 
$e^a{}_i = \pm 1$ for $a =1, 2, 3$ and $i = 1, \dots, n^a$. 
Using the non-commutativity relations Eq.(\ref{noncommutativity}) 
in the zero temperature limit and the decomposition Eqs.(\ref{decomp1}, \ref{decomp2}), we have 
\beq
\vert B \rangle &=& \sum_{n^1, n^2, n^3} \frac{1}{n^1! n^2! n^3!} 
\left(\frac{V}{2} \right)^{n^1+ n^2 +n^3} \int \prod^{n^1}_{i=1} d\s^1{}_i
\prod^{n^2}_{i=1} d\s^2{}_i\prod^{n^3}_{i=1} d\s^3{}_i \nn\\
&& ~~~\exp \left\{ -i \frac{3}{2} \th \sum_{a , d=1}^3 \sum_{b, c =1}^2 \sum_{i=1}^{n^a}
\sum_{\s^a{}_i > \s^b{}_j}^{n^b}
e^a{}_i R_{ab} \,\e_{bc}\,(R^t)_{cd}\, e^d{}_j \right\} \nn\\
&&~~~{\boldsymbol T}\exp \left\{ i \sqrt{\frac{3}{2}} \sum_{a=1}^3 \sum_{b=1}^2 
\sum_{i=1}^{n^a}  R_{ab} Z^b (\s^a{}_i) 
\right\}\vert B_E \rangle 
\eeq
We may rewrite the phase factor, arising from the non-commutativity of the string coordinates 
$X^a$, $a= 1, 2$, as follows
\beq
\frac{3}{2} \th \sum_{a , d=1}^3 \sum_{b, c =1}^2 
\sum_{i=1}^{n^a}\sum_{\s^a{}_i > \s^b{}_j}^{n^b}
e^a{}_i R_{ab} \,\e_{bc}\,(R^t)_{cd}\, e^d{}_j
= \frac{3}{2} \th \sum_{a , b=1}^3 \sum_{i=1}^{n^a}
\sum_{\s^a{}_i > \s^b{}_j}^{n^b}\frac{1}{\sqrt{3}} e^a{}_i (N)_{ab} e^b{}_j,
\eeq
where
\beq
(N) = \left(\begin{array}{rrr} 0 & 1 & ~1 \\ -1 & 0 & ~1 \\ -1 & -1 & ~0 \end{array} \right).
\eeq

Since $e^a{}_i$ and the components of the matrix $(N)$ are only $+1$ or $-1$, two 
different non-commutativity parameters $\th$ and $\widehat\th$ produce the same phase factor if they
satisfy the following condition
\beq
\frac{\sqrt{3}}{2} \th = \frac{\sqrt{3}}{2} \widehat\th + 2\pi n , ~~~ n \in {\bf Z} .\label{noncom1}
\eeq
Two points $(\a,\b)$ and $(\whata,\whatb)$ on the two dimensional parameter space, may 
correspond to the exactly same boundary state and the partition function, if they 
have the same closed string world sheet metric and satisfy the equivalence relation of 
the non-commutativity parameter Eq.(\ref{noncom1})
\begin{subequations}
\beq
\frac{\a}{\a^2+\b^2} &=& \frac{\widehat\a}{\widehat\a^2+ \widehat\b^2} , \label{equiv1}\\
\frac{\b}{\a^2+\b^2} &=& \frac{\widehat\b}{\widehat\a^2+\widehat\b^2} + \frac{2n}{\sqrt{3}} .\label{equiv2}
\eeq
\end{subequations}
These equivalence relation identifies points on the two dimensional parameter space. 
If we define a complex parameter \cite{callan91}
\beq
z = \a+ \b i, ~~~
\eeq
we may rewrite
the closed string metric $G$ and the noncommutativity parameter $\theta$ as 
\beq
G = \frac{|z|^2}{\left({\rm Re}\, z\right)^2} I , ~~~
\theta = 2\pi \, {\rm Im} \left(\frac{1}{z}\right),
\eeq
and the equivalence relation Eqs.(\ref{equiv1}, \ref{equiv2}) succinctly as 
\beq
\frac{1}{\widehat z} = \frac{1}{z} + i \frac{2}{\sqrt{3}} n, ~~~n \in {\bf Z}.
\eeq
We should note that it differs from the equivalence relation of the DHM on a square lattice \cite{callan91}
by the factor of $2/\sqrt{3}$
\beq
\frac{1}{\widehat z} = \frac{1}{z} + i n,~~~n \in {\bf Z}.
\eeq

The $O(2,2;{\bf R})$ transformation between the commutative basis 
$\{\a^I_n, \tilde\a^I_n\}$ and the non-commutative basis $\{\b^I_n, \tilde\b^I_n\}$ 
corresponds
to the T-dual transformation given as
\beq
T = \left(\begin{array}{cc}
  a & b \\
  c & d 
\end{array}\right)
=  \left(\begin{array}{cc}
  I & 0 \\
 \frac{\b}{\a^2+\b^2} \bole  & I
\end{array}\right).
\eeq
Two equivalent DHMs on a triangular lattice are also related by an $O(2,2;{\bf R})$ T-dual transformation
of which explicit expression is given as 
\beq
T^{-1} (\whata, \whatb) T(\a,\b) = \left(\begin{array}{cc}
  I & 0 \\
\left(\frac{\b}{\a^2+ \b^2} - 
\frac{\whatb}{\whata^2+ \whatb^2}\right) \bole  & I 
\end{array}\right) =  \left(\begin{array}{cc}
  I & 0 \\
\frac{2n}{\sqrt{3}} \bole  & I 
\end{array}\right) . \label{exactq}
\eeq
Thus, the subgroup of the $O(2,2;{\bf R})$ T-duality of string theory, which preserves the boundary periodic potential, generated by the T-dual transformation Eq.(\ref{exactq}), is the exact symmetry group of the 
DHM on a triangular lattice in the zero temperature.

\begin{figure}[htbp]
   \begin {center}
    \epsfxsize=0.8\hsize
%
    \epsfbox{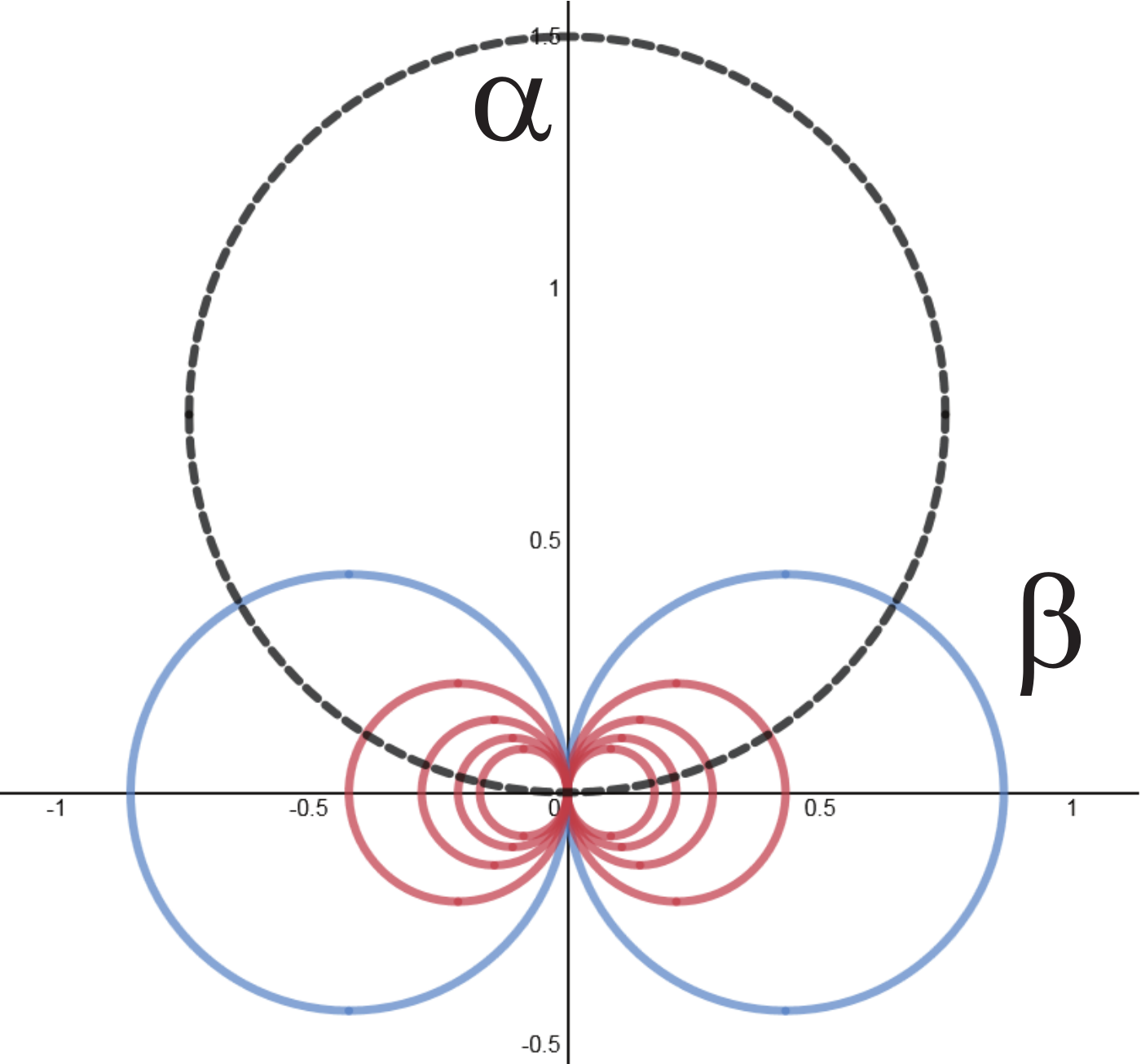}
   \end {center}
   \caption {\label{magic} Magic Circles of The Dissipative Hofstadter Model on a Triangular Lattice.}
\end{figure}

The equivalence relation defines circles on the parameter space. All the points on the 
circle 
\beq
\left(\a - \frac{\sqrt{\det G}}{2} \right)^2 + \b^2 = \left(\frac{\sqrt{\det G}}{2}\right)^2, \label{first}
\eeq
have the same closed string metric $G_{ab} = \sqrt{\det G}\, \delta_{ab}$, 
in the non-commutative basis and all the points on the 
circles 
\beq
\a^2 + \left(\b - \frac{1}{2\left(\th+ 2 n /\sqrt{3}\right)}\right)^2 = 
\left(\frac{1}{2\left(\th+ 2 n /\sqrt{3}\right)}\right)^2, ~~~ n \in {\bf Z} \label{second}
\eeq
share the same non-commutativity parameter $\th$. 
Thus the points where the circles of Eq.(\ref{first}) and Eq.(\ref{second}) meet together 
are equivalent to each other. 
Especially when $\th =0$, the models corresponding to the points on the circles Eq.(\ref{second}) are equivalent to
the boundary sine-Gordon model on a triangular lattice. These circles are 
termed as magic circles \cite{callan91,Callan:1995ck} (Fig.\ref{magic}.)
\beq
\a^2 + \left(\b-\frac{1}{4 n / \sqrt{3}} \right)^2 = \frac{3}{16 n^2}, ~~~~n \in {\bf Z} . 
\eeq

The renormalization group (RG) exponent of the boundary interaction, hence the critical behavior of 
the model is determined by $\det G$. It may need a lengthy perturbation theory 
analysis to fix the RG exponents of the DHM on a triangular lattice, which depend 
on details of the perturbation theory. Embedding the DHM on a triangular lattice in a 
three dimensional model, which requires three string coordinate fields, leads us to the 
following critical circle (the dotted circle in Fig.\ref{magic})
\beq
\left(\a - \frac{3}{4} \right)^2 + \b^2 = \frac{9}{16}.
\eeq
On the critical circle the periodic boundary interaction may be represented by fermion bilinear operators. 
On the points where the critical circle coincides with the magic circles, the DHM becomes exactly solvable 
in terms of free fermion fields. These points are called magic points \cite{Callan:1995ck,Lee;2007}. 


\section{Conclusions}

Dualities are very important to understand the critical behaviors of low dimensional quantum systems, 
since the global structures of the quantum systems may be determined by them.
The exact $O(2,2;R)$ duality of the DHM on a square lattice has been useful to 
study the structure of the phase diagram of the model. The duality also offers two equivalent 
descriptions of the model: the model in the commutative basis and the model in the non-commutative 
basis. In this context, the non-commutative string theory has been proven to be a valuable tool to 
analyze the DHM on a square lattice 
\cite{callan91,Callan:1992vy,freed93,Callan:1995ck,Lee2009,Leequon,Lee;2007}. 
In the present paper, we extend the previous works on a
square lattice to the DHM on a triangular lattice. The model itself is an interesting one,
since it corresponds to a model of quantum Brownian motion on a triangular lattice in the 
presence of uniform magnetic field. The model is also important to study the junctions of 
quantum wires, as we have shown that the model describes the junction of three quantum wires in
the presence of uniform magnetic field. The DHM on a square lattice only depicts the junction of 
two wires. 

A fermion model of the junction of three quantum wires has been discussed in refs.\cite{chamon,oshikawa2006},
by mapping the model onto a DHM on a triangular lattice. Bosonizing the model of the junction of three
wires, they encountered a non-trivial Klein factors in the boson theory. The main reason to map the 
fermion model onto the DHM on a triangular lattice was to replace the phases due to the Klein factors 
by the phases of the non-commutativity between the boson fields
induced by the interaction with the uniform magnetic field. 
Even in the absence of the magnetic field, the fermion model is 
mapped onto the boson model of DHM on a triangular 
lattice. However, it is also possible to transcribe the fermion model in the absence of 
the magnetic field onto
a boson model, which does not contain any non-trivial Klein factor, {\it i. e.}, 
the boundary sine-Gordon model on a triangular lattice, if the representations of the Klein factors of 
the fermion fields are judiciously chosen \cite{Lee2009q,Leeklein}. 
Different mappings of the model may result in different phase diagrams. We have not 
yet given a perturbation analysis on the RG flow of the periodic potential operator 
on a triangular lattice. The RG exponent of the operator may 
differ from the naive scale dimension if a non-trivial interaction is present. 
It may depend on details of the perturbation theory. 
We will discuss the perturbation analysis of the DHM on a triangular lattice elsewhere in a separate 
paper. The exact duality discussed in the present work may help us to understand the 
critical behaviors of the DHM on a triangular lattice, hence the junction of three quantum wires.

\vskip 1cm

\noindent{\bf Acknowledgments}\\
This work was supported by Kangwon National University.


%







\end{document}